\documentclass[12pt]{article}

\usepackage{amsfonts}
\usepackage{amsmath}
\newtheorem{theorem}{\bf{Theorem}}[section]
\newtheorem{lemma}{\bf{Lemma}}[section]

\numberwithin{equation}{section}

\def \ben{\begin{eqnarray*}}
\def \een{\end{eqnarray*}}
\def \bea{\begin{eqnarray}}
\def \eea{\end{eqnarray}}

\begin{document}
\baselineskip 27pt
\begin{center}
\small{Smoothed and Iterated Bootstrap Confidence Regions for Parameter Vectors}\\

Santu Ghosh and Alan M. Polansky\\

Wayne State University and Northern Illinois University, USA\\

\end{center}

\begin{abstract}
The construction of confidence regions for parameter vectors is a difficult problem in the nonparametric setting, particularly when the sample size is not large. The bootstrap has shown promise in solving this problem, but empirical evidence often indicates that some bootstrap methods have difficulty in maintaining the correct coverage probability, while other methods may be unstable, often resulting in very large confidence regions. One way to improve the performance of a bootstrap confidence region is to restrict the shape of the region in such a way that the error term of an expansion is as small an order as possible. To some extent, this can be achieved by using the bootstrap to construct an ellipsoidal confidence region. This paper studies the effect of using the smoothed and iterated bootstrap methods to construct an ellipsoidal confidence region for a  parameter vector. The smoothed estimate  is based on a multivariate kernel density estimator. This paper establishes a bandwidth matrix for the smoothed bootstrap procedure that reduces the asymptotic coverage error of the bootstrap percentile method ellipsoidal confidence region.  We also provide an analytical adjustment to the nominal level to reduce the computational cost of the iterated bootstrap method. Simulations demonstrate that the methods can be successfully applied in practice.

\end{abstract}
keywords: Bandwidth Matrix, Bootstrap Percentile Method, Bootstrap\\*Percentile-$t$ Method, Iterated Bootstrap Method, Edgeworth Expansion,\\* Smooth Function Model.

\section{Introduction} 
\numberwithin{equation}{section}
The construction of bootstrap confidence intervals has been studied  extensively over the past few decades. Early criticism of the bootstrap percentile method (Efron, 1979) led to several improvements of the methodology, including the bias corrected method (Efron, 1981), the bias-corrected and accelerated method (Efron, 1987), and the studentized method (Efron, 1982). Methods based on pre-pivoting, the iterated bootstrap, and calibration were developed by Beran (1987), Hall (1986), and Loh (1987). Hall (1988) provided a systematic method for comparing confidence intervals based on Edgeworth expansion theory. Implementation of the smoothed bootstrap with the specific purpose of improving the coverage properties of confidence intervals has been discussed by Guerra, Polansky and Schucany (1997), Polansky and Schucany (1997), and Polansky (2001). However,  multivariate confidence regions have received limited consideration and it is difficult to extend most of the existing univariate procedures directly to the multivariate case.

Let $X_1,X_2,\ldots,X_n$ be a set of independent and identically distributed $p$-dimensional random vectors following a distribution $F$. Let $\theta=t(F)$ be a parameter vector and $\hat{\theta}_n$ is a plug-in estimator of $\theta$ and $\hat{\Omega}_n$ is a consistent estimator of the asymptotic covariance matrix $\Omega$ of $n^{1/2}\hat{\theta}_n$, assume that $\Omega$ is non-singular. Then a $100\alpha\%$ confidence region for $\theta$ has the form
\[
\mathcal{R}=
\{\hat{\theta}_n-n^{-1/2}\hat{\Omega}_n^{1/2}r:r\in\mathcal{R}_\alpha\},
\]
\noindent
where 
$\mathcal{R}_\alpha\subset\mathbb{R}^{d}$ 
is any region such that $P[\sqrt{n}\hat{\Omega}_n^{-1/2}(\hat{\theta}_n-\theta)\in\mathcal{R}_\alpha]=\alpha$. The shape of the region $\mathcal{R}$ depends on the shape of the region $\mathcal{R}_\alpha$. In this paper we concentrate on ellipsoidal confidence regions, which are generalizations of univariate symmetric confidence intervals. In particular, if $\mathcal{R}_\alpha$ is a $d$-variate sphere centered at origin, then  $\mathcal{R}$ becomes an ellipsoidal confidence region.  In practice the bootstrap is often used to estimate $\mathcal{R}_\alpha$. 

A simpler method for computing an ellipsoidal confidence region for $\theta$ is based on extending the bootstrap percentile method of Efron (1979) to the multivariate case.  Let $\mathcal{R}_{BP}$ be a bootstrap percentile method ellipsoidal confidence region for $\theta$. For a given nominal level $\alpha$,  we shall prove that  
\bea \label{eq:a1}
P(\theta\in \mathcal{R}_{BP})= \alpha +n^{-1}Q(\chi^{2}_{v,\alpha})+O(n^{-2}),
\eea
 where $Q(\chi^{2}_{v,\alpha})$ is a polynomial in $\chi^{2}_{v,\alpha}$. The coefficients of $Q(\chi^{2}_{v,\alpha})$ are functions of population moments  and $\chi^{2}_{v,\alpha}$ is  the $\alpha$ quantile  of an chi-square distribution with $v$ degrees  of freedom. Equation (\ref{eq:a1}) shows that $\mathcal{R}_{BP}$ is  second order accurate. An alternative method for constructing ellipsoidal confidence region for $\theta$ is the bootstrap percentile-$t$ method.  From an asymptotic viewpoint, the bootstrap percentile-$t$ method is fourth-order accurate, see  Hall (1992, Section 4.2). Our empirical studies show that while the bootstrap percentile-$t$ method has acceptable coverage probabilities, it can be unstable and can produce large ellipsoidal confidence regions when the sample size is small. 

The natural idea is to improve the coverage probability of $\mathcal{R}_{BP}$. In the univariate setup, smoothed and iterated bootstrap methods have potential application in the construction of confidence intervals. Both of these methods are easily implementable as practical procedures for routine use. To our knowledge, so far the use of the smoothed and iterated bootstrap methods have been not been explored in the case of multivariate regions.  To improve the coverage probability of $\mathcal{R}_{BP}$, we consider a multivariate version of the smoothed and iterated bootstrap methods. However, the performance of the smoothed bootstrap heavily depends on the choice of the bandwidth matrix and the latter  method is computationally expensive, specifically in the multivariate case. In this paper our contribution are, $(i)$ we establish an explicit form of the bandwidth matrix which is succeed in reducing the order of coverage error of $\mathcal{R}_{BP}$ to  $O(n^{-2})$  and $(ii)$ we provide an analytical correction is  to the nominal level to avoid the double bootstrap for constructing the iterated bootstrap percentile method ellipsoidal confidence region. We also show that the resulted region also reduces the coverage error of $\mathcal{R}_{BP}$ to $O(n^{-2})$.         

The remainder of the paper is organized as follows. Section 2 introduces the smoothed and iterated bootstrap methods in the case of a mean vector.  Section 3 extends  these methods for a  multivariate smooth function of a mean vector. Simulation results are reported in Section 4. Section 5 concludes and Appendix A contains some technical details.

\section{Bootstrap confidence regions for a Mean vector}
\numberwithin{equation}{section}

Let $\theta=E_{F}(X_{n})$ be a mean vector of $F$ and  assume that the covariance matrix $\Sigma$, of $F$, is positive definite and unknown. We are interested in constructing an ellipsoidal confidence region for $\theta$.
Let 
\[
\hat{\theta}_{n}=n^{-1}\sum^{n}_{i=1}X_i,
\] 
and
 \[
 \hat{\Sigma}_n=n^{-1}\sum_{i=1}^n(X_i-\bar{X}_n)(X_i-\bar{X}_n)'.
 \]

To facilitate our discussion of the bootstrap percentile method ellipsoidal confidence region, let $X^*_1,\ldots,X^*_n$ be a random sample from the empirical distribution $\hat{F}_{n}$. 
Let 
\[
\hat{\theta}^*_n=n^{-1}\sum^n_{i=1} X^*_i.
\] 

and
\[
\hat{\Sigma}^*_n=
n^{-1}\sum_{i=1}^n(X^*_i-\bar{X}^*_n)(X^*_i-\bar{X}^*_n)'.
\]
The bootstrap percentile ellipsoidal method confidence region for $\theta$ with approximate coverage probability $\alpha$ has the form
\[
\mathcal{R}_{BP}=
\{\hat{\theta}_n-n^{-1/2}\hat{\Sigma}^{1/2}_ns:s\in\mathcal{S}_{BP}\}, 
\]
where $\mathcal{S}_{BP}$ denotes a $p$-variate sphere centered at origin such that $P^*(S^*\in\mathcal{S}_{BP})=\alpha$ and $S^*=\sqrt{n}\hat{\Sigma}^{-1/2}_{n}(\hat{\theta}^{*}_{n}-\hat{\theta}_{n})$. $P^{*}$ denotes the probability measure  conditional on $X_1,\ldots, X_n$.

An alternative method is the \textit{bootstrap percentile-$t$ ellipsoidal confidence region} for $\theta$ with approximate coverage probability $\alpha$, given by
\[
\mathcal{R}_{BT}=
\{\hat{\theta}_n-n^{-1/2}\hat{\Sigma}^{1/2}_ns:s\in\mathcal{S}_{BT}\}, 
\]
where $\mathcal{S}_{BT}$ denotes a $p$-variate sphere centered at the origin such 
that $P^*(U^*\in\mathcal{S}_{BT})=\alpha$ and $U^*=\sqrt{n}\hat{\Sigma}^{* -1/2}_n(\hat{\theta}^*_n-\hat{\theta}_n)$.  $\mathcal{R}_{BT}$ can be unstable  if there is a significant conditional probability under $\hat{F}_{n}$ that $\hat{\Sigma}^*_n$ is nearly singular. We begin with the asymptotic expansion for the coverage probability of $\mathcal{R}_{BP}$. 

The following assumptions are made throughout this section:
\begin{enumerate} 
\item 
The distribution $G$ of $\left[
\begin{matrix}
vec(X) \\ vech(X X^{'})
\end{matrix}
\right]$ 
satisfies the multivariate version of the Cram\'{e}r continuity condition. The condition holds provided $G$ has a non-degenerate absolutely continuous component. See Hall (1992, Pages 66-67). 
\item Assume all moments of order 6 of $Y$ are finite. 
That is $E(\|Y\|^{6})<\infty$.
\end{enumerate}

 The assumptions 1--2 guarantee that the $S=\sqrt{n}\Sigma^{-1/2}(\hat{\theta}_n-\theta)$ and $U=\sqrt{n}\hat{\Sigma}^{-1/2}_n (\hat{\theta}_n-\theta)$ have  three-term Edgeworth expansions. That is, 
\ben
\sup_{B}\left|P(S\in B)-\int_{B}\gamma_{3,n}(s)ds\right|= O(n^{-2}),
\een 
and
\ben
\sup_{B}\left|P(U\in B)-\int_{B}\eta_{3,n}(u)du\right|=O(n^{-2}),
\een
\noindent
where  the supremum is taken over the sets $B \in{\mathbb{R}}^{d}$, which are unions of a finite number of convex sets and $\gamma_{3,n}(s)=\phi_{p}(s)[1+\sum^{3}_{i=1}n^{-i/2}Q_{i}(s)]$ and $\eta_{3,n}(u)=\phi_{p}(u)[1+\sum^{3}_{i=1}n^{-i/2}R_{i}(u)]$. $\phi_{p}(.)$ is the density function of the $p$-variate standard normal distribution and  functions $Q_{i}(s)$ and $R_{i}(u)$  are  polynomials of degree $3i$ and are odd(even) polynomials for odd(even) $i$, respectively. See Barndorff-Nielsen and Cox (1989, Chapter 5) and Hall (1992, Section 4.2). 

\begin{theorem}\label{T:Th1} Under Assumptions 1 and 2, 
\begin{equation}\label{eq:a2}
P(\theta\in\mathcal{R}_{BP})=
\alpha+n^{-1}\left[q_1(\chi^2_{p,\alpha})-q_2(\chi^2_{p,\alpha})\right]
g_p(\chi^{2}_{p,\alpha})+O(n^{-2}). 
\end{equation}
\end{theorem}
In general, $q_1(\chi^2_{p,\alpha})$ differs from $q_2(\chi^2_{p,\alpha})$
and therefore an immediate consequence of Equation (\ref{eq:a2}) is that $\mathcal{R}_{BP}$ is second-order accurate. Where
 
\begin{eqnarray}\label{eq:a3} 
q_{1}(\chi^{2}_{p,\alpha})&=&\left[\frac{\kappa^{(2)}_{3}}{8}+
\frac{\kappa^{(1)}_{3}}{12}-\frac{\kappa^{(1)}_{4}}{8}\right]\frac{2}{p}
\chi^{2}_{p,\alpha}\nonumber\\
&&+\left[\frac{\kappa^{(1)}_4}{8}-\frac{\kappa^{(2)}_3}{4}-
\frac{\kappa^{(1)}_3}{6}\right]\frac{2}{p}\frac{(\chi^{2}_{p,\alpha})^2}{p
+2}\nonumber\\
&&+\left[\frac{\kappa^{(2)}_{3}}{8}+\frac{\kappa^{(1)}_{3}}{12}\right]
\frac{2}{p}\frac{(\chi^{2}_{p,\alpha})^{3}}{(p+2)(p+4)}. 
\end{eqnarray}
and

\begin{eqnarray} \label{eq:a4} 
q_{2}(\chi^{2}_{p,\alpha})&=&\left[\frac{p}{4}(p+2)+
\frac{\kappa^{(1)}_{4}}{2}-\frac{\kappa^{(1)}_{3}}{6}\right]
\frac{2}{p}\chi^{2}_{p,\alpha}\nonumber\\
&&+\left[\frac{p}{4}(p+2)+\frac{\kappa^{(1)}_{3}}{3}-
\frac{\kappa^{(1)}_{4}}{4}\right]\frac{2}{p}
\frac{(\chi^{2}_{p,\alpha})^{2}}{p+2}\nonumber\\
&&+\left[\frac{\kappa^{(1)}_{3}}{3}+
\frac{\kappa^{(2)}_{3}}{2}\right]\frac{2}{p}\frac{(\chi^{2}_{p,\alpha})^{3}}{(p+2)(p+4)}, 
\end{eqnarray}
where $\kappa^{(1)}_{3}$ and $\kappa^{(1)}_{4}$ are the measures of the multivariate skewness and kurtosis of $F$ introduced by Mardia (1970), and  $\kappa^{(2)}_{3}$ is the measure of multivariate skewness introduced by Isogai (1983). $g_p(\chi^{2}_{p,\alpha})$ is the density function of a chi-square random variable with $p$ degrees of freedom.  The functions $q_{1}(\chi^{2}_{p,\alpha})$ and $q_{2}(\chi^{2}_{p,\alpha})$ appear in asymptotic expansions of square radii of the $p$-variate spheres $\mathcal{S}_{S}$ and $\mathcal{S}_{U}$,
\[r^{2}_{S}=\chi^{2}_{p,\alpha}+n^{-1}q_{1}(\chi^{2}_{p,\alpha})+O(n^{-2}),\] and

\[r^{2}_{U}=\chi^{2}_{p,\alpha}+n^{-1}q_{2}(\chi^{2}_{p,\alpha})+O(n^{-2}),\] 
\noindent
where $\mathcal{S}_{S}$ and $\mathcal{S}_{U}$ are such that $P(S\in \mathcal{S}_{S})=\alpha$ and $P(U\in \mathcal{S}_{U})=\alpha$. Proofs of these expansions are given in the appendix.

To improve the coverage accuracy of $\mathcal{R}_{BP}$, our first approach will be to apply the smoothed bootstrap to $\mathcal{R}_{BP}$.

\subsection{The Smoothed Bootstrap}
\numberwithin{equation}{section}
When a smoothed estimate of $F$ is used to calculate a bootstrap estimate, then the process is known as smoothed bootstrap. Let $\hat{F}_{n,H_{n}}$ denote a smoothed version of $\hat{F}_{n}$ based on a $p$-dimension kernel density estimator. Let $\hat{f}_{n,H_{n}}$ be the density corresponding to $\hat{F}_{n,H_{n}}$, which has the following form
\[
\hat{f}_{n,H_{n}}(x)= 
n^{-1}|H_{n}|^{-1/2}\sum^{n}_{i=1}K[H^{-1/2}_{n}(x-X_{i})].
\]
The matrix $H_{n}$ is a $p\times p$ positive definite matrix is called a \textit{bandwidth matrix}, which is usually function of the sample size $n$.  Here, we assume that $K$ is the standard $p$-variate normal density function because it has a very specific cumulant structure which we will use to our advantage. We assume that  
$H_n=O(n^{-k})$ for some $k>0$, the value of $k$ will  be discussed later. That is, $h_{ij,n}=O(n^{-k})$, where $h_{ij,n}$ is the $(i,j)$-th element of $H_n$. See Wand and Jones (1995) for a discussion of the  general theory of multivariate kernel density estimators.

Another alternative to the bootstrap percentile method is based on the smoothed 
bootstrap.  Let $Y^*_1,\ldots,Y^*_n$ be a random sample from $\hat{F}_{n, H_{n}}$. In practice we can simulate $Y^*_1,\ldots,Y^*_n$ based on the fact that $Y^{*}_{i}$ has the same conditional distribution as $X^{*}_{i}+W_{i}$, where $X^{*}_{i}$ and $W_{i}$ are independent random vectors follow $\hat{F}_{n}$ and $N_{p}(0, H_{n})$, respectively. Then the smoothed bootstrap percentile method ellipsoidal confidence region for $\theta$ with approximate coverage probability $\alpha$ is
\[
\mathcal{R}_{SBP}=
\{\hat{\theta}_n-n^{-1/2}\tilde{\Sigma}^{1/2}_ns:s\in\mathcal{S}_{SBP}\}, 
\]
where 
$\mathcal{S}_{SBP}$ denotes a $p$-variate sphere centered at the origin such that 
$P^*(\tilde{S}^*\in\mathcal{S}_{SBP})=\alpha$, where $\tilde{S}^*=\sqrt{n}\tilde{\Sigma}^{-1/2}_n(\tilde{\theta}^*_n-\hat{\theta}_n)$, $\tilde{\Sigma}_n=\hat{\Sigma}_n+H_n$, and $\tilde{\theta}^*_n$ is the version of $\hat{\theta}^*_n$, based on $Y^*_1,\ldots,Y^*_n$. 

The following result establishes the effect that smoothing has on the bootstrap percentile method ellipsoidal confidence region. In particular, it shows that using an appropriate choice of the bandwidth matrix $H_n$,  $\mathcal{R}_{SBP}$ is fourth-order accurate.

\begin{theorem}\label{T:Th2} 
If 
$H_n=(n\chi^2_{p,\alpha})^{-1}
\{q_2(\chi^2_{p,\alpha})-q_1(\chi^2_{p,\alpha})\}\Sigma$ and 
$q_2(\chi^2_{p,\alpha})-q_1(\chi^2_{p,\alpha})> 0$, then under 
Assumptions 1--2, 
$\mathcal{R}_{SBP}$ is fourth order-accurate. 
That is $P(\theta\in \mathcal{R}_{SBP})=\alpha+O(n^{-2})$. 
\end{theorem}
One can observe from the Theorem \ref{T:Th2} that the smoothed bootstrap reduces the order of the asymptotic coverage error of the bootstrap percentile method from $O(n^{-1})$ to $O(n^{-2})$. In other words, the smoothed bootstrap percentile method is as asymptotically accurate as the bootstrap percentile-$t$ method. The condition that $q_2(\chi^2_{p,\alpha})-q_1(\chi^2_{p,\alpha})> 0$ is sufficient in that a reduction in the asymptotic coverage error is not possible unless the condition holds. This condition is closely related to how smoothing allows the correction to take place. One can observe from the expansion in Equation (\ref{eq:a2}) that when the condition holds, the confidence region asymptotically has an \textit{under-coverage problem}. Smoothing adds variation to the resampled values of the sample mean, which in turn increases the area of the corresponding confidence region so that the boundaries of the smoothed region coincide better with the theoretical boundary for the studentized method.  The smoothing method can not be applied to the problems that asymptotically over-cover, without significant modification. 

It is easy to observe that the optimal bandwidth matrix for the smoothed bootstrap depends on unknown parameters and that to apply the smoothed bootstrap in practice we have to replace $H_n$ by an estimator $\hat{H}_n$. Because $q_2(\chi^2_{p,\alpha})-q_1(\chi^2_{p,\alpha})$ is a function of the population moments, the most direct method for estimating $H_n$ is to use a plug-in bandwidth matrix, where we replace the population moments in $\hat{H}_n$ with sample moments. Thus, a simple plug-in estimator of $H_n$ is given by 
\[
\hat{H}_n=(n\chi^2_{p,\alpha})^{-1}\{\hat{q}_2(\chi^2_{p,\alpha})-
\hat{q}_1(\chi^2_{p,\alpha})\}\hat{\Sigma}_n,
\]  
provided $\hat{q}_2(\chi^2_{p,\alpha})-\hat{q}_1(\chi^{2}_{p,\alpha})> 0$. Because $\hat{q}_1(\chi^2_{p,\alpha})$ and $\hat{q}_2(\chi^2_{p,\alpha})$ are linear combinations of sample moments, it follows that $\hat{q}_1(\chi^2_{p,\alpha})=q_1(\chi^2_{p,\alpha})+O_p(n^{-1/2})$ and $\hat{q}_2(\chi^2_{p,\alpha})=q_2(\chi^{2}_{p,\alpha})+O_p(n^{-1/2})$. Similarly $\hat{\Sigma}_n=\Sigma+O_p(n^{-1/2})$. Combining these results yields $\hat{H}_n=H_n+O_p(n^{-3/2})$. In result below we show that this plug-in estimator $\hat{H}_{n}$ is accurate enough to insure the fourth-order accuracy of the smoothed bootstrap percentile method.
\begin{theorem}\label{T:Th3} 
If $\hat{H}_n=(n\chi^2_{p,\alpha})^{-1}
\{\hat{q}_2(\chi^2_{p,\alpha})-
\hat{q}_1(\chi^2_{p,\alpha})\}\hat{\Sigma}_n$ 
and
$P[\hat{q}_2(\chi^2_{p,\alpha})-\hat{q}_1(\chi^2_{p,\alpha})>0]=1$ as 
$n\rightarrow{\infty}$ then 
$P(\theta\in\hat{\mathcal{R}}_{SBP})=\alpha+O(n^{-2})$,  
where $\hat{\mathcal{R}}_{SBP}$ is the region $\mathcal{R}_{SBP}$ using 
the estimated bandwidth matrix $\hat{H}_n$. 
\end{theorem}

The assumption involving $\hat{q}_2(\chi^2_{p,\alpha})-\hat{q}_1(\chi^2_{p,\alpha})$ in Theorem \ref{T:Th3} holds if $\hat{\mathcal{R}}_{BP}$ has asymptotically under coverage problem. In practice $\hat{\mathcal{R}}_{SBP}$ can not be constructed  when the empirical bandwidth matrix $\hat{H}_n$ is negative. As mentioned before, this is problem is most significant if $\mathcal{R}_{BP}$ has an asymptotically over coverage problem. Hence, in such cases, a reduction may be required in the variance of $\hat{\theta}_n^{*}$. Therefore, the smoothed bootstrap percentile method ellipsoidal confidence region can be constructed as   
\[
\hat{\mathcal{R}}_{SBP}^{1}=
\{\hat{\theta}_n-n^{-1/2}\tilde{\Sigma}^{1/2}_{n,1}s:s\in\mathcal{S}_{SBP}^{1}\}, 
\]
where
\[ \tilde{\Sigma}_{n,1}= \hat{\Sigma}_n -\hat{H}_{n,1},\]
where $\hat{H}_{n,1}=(n\chi^2_{p,\alpha})^{-1}
\{\hat{q}_1(\chi^2_{p,\alpha})-\hat{q}_2(\chi^2_{p,\alpha})\}\hat{\Sigma}_n$ and 
$\mathcal{S}_{SBP}^{1}$ is the version of $\mathcal{S}_{SBP}$ using $\tilde{\Sigma}_{n,1}$.   Additionally, it can be shown that $P(\theta\in\hat{\mathcal{R}}_{SBP}^{1})=\alpha+O(n^{-2})$. Therefore, depends on the sign of $\hat{H}_n$, we can either construct $\hat{\mathcal{R}}_{SBP}$ or $\hat{\mathcal{R}}_{SBP}^{1}$.

In the next subsection we investigate the asymptotic effects on coverage error of calibrating the nominal coverage level of $\mathcal{R}_{BP}$.

\subsection{The Iterated Bootstrap Method}
\numberwithin{equation}{section} 

Iterated bootstrap shows assurance to improve coverage probabilities for constructing confidence regions. To construct an iterated bootstrap confidence region we usually make an additive correction to the nominal coverage level. The additive correction term is determined using a computationally expensive Monte Carlo simulation method that involves the double bootstrap. We will provide an analytical correction to nominal level of $\mathcal{R}_{BP}$ and this correction replaces the need of second-level bootstrapping of the iterated bootstrap method.

To facilitate the discussion of the iterated bootstrap  percentile method ellipsoidal confidence region, let $\mathcal{X}^{*}$ denote a generic first level bootstrap sample drawn randomly, with replacement, from $\hat{F}_{n}$ and similarly $\mathcal{X}^{**}$ denotes a generic second level bootstrap sample drawn randomly, with replacement, from 
$\hat{F}_{n}^{*}$, where $\hat{F}_{n}^{*}$ is the empirical distribution function based on $\mathcal{X}^{*}$. Let $S^{**}=\sqrt{n}\hat{\Sigma}^{*-1/2}_{n}(\hat{\theta}^{**}_{n}-\hat{\theta}^{*}_{n})$ denote the version of $S^{*}$ based on $\hat{F}_{n}^{*}$. Then we define the theoretical iterated bootstrap  percentile method ellipsoidal confidence region for $\theta$ to be
\[
\mathcal{R}_{RBP}=\{\hat{\theta}_{n}-n^{-1/2}\hat{\Sigma}^{1/2}_{n}s:s\in \mathcal{S}_{BP,\alpha+u_n}\}, 
\] 
\noindent  
where $u_n$ satisfies
\[ P[\hat{\theta}_{n}\in \{\hat{\theta}^{**}_{n}-n^{-1/2}\hat{\Sigma}^{*1/2}_{n}s:s\in \mathcal{S}^{*}_{BP,\alpha+u_n}\}|\mathcal{X}, \mathcal{X}^*]=\alpha.\]
\noindent
In practice, the confidence region $\mathcal{R}_{RBP}$ is constructed using the double bootstrap. Here we provide an analytical approximation for $u_n$ to avoid the need of the double bootstrap. 

Under the assumptions 1--2, it can be easily shown that $u_{n}$ has the following expansion
\bea\label{eq:a5}
u_n=n^{-1} [\hat{q}_{2}(\chi^2_{p,\alpha})-\hat{q}_{1}(\chi^2_{p,\alpha})]g_{p}(\chi^2_{p,\alpha}) +O_{p}(n^{-2}). 
\eea 

Define
$\tilde{u}_{n}= n^{-1} [\hat{q}_{2}(\chi^2_{p,\alpha})-\hat{q}_{1}(\chi^2_{p,\alpha})]g_{p}(\chi^2_{p,\alpha}),$   by this means we construct
\[
\mathcal{R}_{AN}=\{\hat{\theta}_{n}-n^{-1/2}\hat{\Sigma}^{1/2}_{n}s:s\in \mathcal{S}_{BP,\alpha+\tilde{u}_{n}}\},\] 
where $AN$ stands for \textit{analytic} due to $\tilde{u}_{n}$.  The following theorem establishes that $\mathcal{R}_{AN}$ has coverage error of order $O(n^{-2})$.

\begin{theorem}\label{T:Th4} Under Assumptions 1--2, $P(\theta \in \mathcal{R}_{AN})=\alpha+O(n^{-2})$.
\end{theorem} 
Theorem \ref{T:Th4} shows that $\mathcal{R}_{AN}$ is fourth-order accurate. The inner level resmapling for $\mathcal{R}_{RBP}$ is avoided by use of $\tilde{u}_{n}$. Therefore, $\mathcal{R}_{AN}$ is  computationally attractive. 

The quantity $\tilde{u}_{n}$ is crucial for construction the region $\mathcal{R}_{AN}$.  For example,  if $\alpha+\tilde{u}_{n} > 1$ for a given sample, the region $\mathcal{R}_{AN}$ is undefined. To overcome such situations, in practice we can use $\alpha^{'}=\max\{\alpha,\min\{1, \alpha+\tilde{u}_{n}\}\}$ instead of $\alpha+\tilde{u}_{n}$.

In the next section we extend the smoothed and iterated bootstrap methods to a multivariate smoothed function of a vector mean.  

\section{Functions of  Mean vectors}
\numberwithin{equation}{section}

Let $g_{1},\ldots,g_{d_{1}}$ be real valued functions on $\mathbb{R}^{p}$. Define $Z_{i}=[g_{1}(X_{i}),\ldots,g_{d_{1}}(X_{i})]^{'},$ $i=1,\ldots,n$. Let $\eta=E(Z_{i})=[Eg_{1}(X_{i}),\ldots,Eg_{d_{1}}(X_{i})]^{'}$  be the mean vector $Z_{i}$, for $i=1,\ldots,n$. Assume, the parameter vector $\theta=A(\eta)$ is defined in terms of the `smoothed function model' (e.g.  Hall (1992), page 52),  where $A=(A_{1},\ldots,A_{d})$ is a Borel measurable function that maps $\mathbb{R}^{d_{1}}$ to $\mathbb{R}^{d}$. Let $\hat{\theta}_{n}=A(\bar{Z}_{n})$ be a plug-in estimator of $\theta$, where \[\bar{Z}_{n}=n^{-1}\sum_{i=1}^{n}Z_{i}.\] 
Assume that $\Omega$ is the asymptotic covariance matrix of $n^{1/2}\hat{\theta}_n$ which can be represented as $C(\eta)\Psi C(\eta)^{'}=h(\eta)$, where $\Psi$ is the covariance matrix of $Z_1$, $C(\eta)=\bigtriangledown A(\eta)$ is the gradient matrix of $A$ at $\eta$. 

Many vector parameters of interest in modern statistics can be studied under a this type of smooth function model. For example, $\theta$ could represent a vector of mean and variance, a vector of correlations. In this case the bootstrap percentile method ellipsoidal confidence region $\tilde{\mathcal{R}}_{BP}$ for $\theta$ is
\[
\tilde{\mathcal{R}}_{BP}=
\{\hat{\theta}_n-n^{-1/2}\tilde{\Omega}^{1/2}_n s:s\in\tilde{\mathcal{S}}_{BP}\}, 
\]
where $\tilde{\mathcal{S}}_{BP}$ denotes a $d$-variate sphere centered at the origin based on the bootstrap percentile method. To avoid the confusion with the mean case, we use $\tilde{\mathcal{R}}_{BP}$ to denote the bootstrap percentile method ellipsoidal confidence region for $\theta$. The coverage probability of $\tilde{\mathcal{R}}_{BP}$ also enjoys similar expansion as in Equation (\ref{eq:a2}) and an immediate consequence  is that $\tilde{\mathcal{R}}_{BP}$  is second-order accurate.       
As we will show in this section, the smoothed and iterated bootstrap procedures that we propose on the inference of a mean vector can also be applied to problems that fit within this more general model as well. 

\subsection{Smoothed Bootstrap}
\numberwithin{equation}{section}

To reduce the complexity of estimating $\theta$ using the smoothed bootstrap, our smoothed bootstrap approach is based on  $Z_1,\ldots,Z_n$. Let $\tilde{f}_{n,H_n}$ be a kernel density estimator of the density function of $Z_1$ with the multivariate standard normal kernel function and we asume that $H_{n}=O(n^{-k})$ for some $k>0$. Let $\tilde{\eta}_n$ and $\tilde{\Psi}_n$ be estimators of $\eta$ and $\Psi$, based on $\tilde{f}_{n,H_n}$. Then $\tilde{\eta}_n=\bar{Z}_{n}$ and $\tilde{\Psi}_n=\hat{\Psi}_n+H_n$, where \[ \hat{\Psi}_n=n^{-1}\sum_{i=1}^n(Z_i-\bar{Z}_n)(Z_i-\bar{Z}_n)'.\] The smoothed bootstrap estimates of $\theta$ and $\Omega$ are given by
\[ \tilde{\theta}_n=A(\tilde{\theta}_n)=A(\bar{Z}_{n})=\hat{\theta}_n\]
and
\[\tilde{\Omega}= C(\tilde{\eta}_n)\tilde{\Psi}_n C(\tilde{\eta}_n)^{'}=C(\bar{Z}_n)(\hat{\Psi}_n +H_n)C(\bar{Z}_n)^{'}=\hat{\Omega}_n+D_n,\]
where $D_n=C(\bar{Z}_n)H_n C(\bar{Z}_n)^{'}.$ 

The smoothed bootstrap percentile method ellipsoidal confidence region for $\theta$ with approximate coverage probability $\alpha$ is given 

\[
\tilde{\mathcal{R}}_{SBP}=
\{\hat{\theta}_n-n^{-1/2}\tilde{\Omega}^{1/2}_n s:s\in\tilde{\mathcal{S}}_{SBP}\}, 
\]
where $\tilde{\mathcal{S}}_{SBP}$ denotes a $d$-variate sphere centered at the origin such that 
$P^*(\tilde{S}^*_n\in\tilde{\mathcal{S}}_{SBP})=\alpha$, where $\tilde{S}^*_n=\sqrt{n}\tilde{\Omega}^{-1/2}_n(\hat{\theta}^*_n-\hat{\theta}_n)$. In next result we establish explicit form of the bandwidth matrix $H_n$ which guarantees that $\tilde{\mathcal{R}}_{SBP}$ is fourth order accurate.

\begin{theorem}\label{T:Th5} 
If 
$H_n=(n\chi^2_{d,\alpha})^{-1}\{
\tilde{q}_2(\chi^2_{d,\alpha})-\tilde{q}_1(\chi^2_{d,\alpha})\}\Psi$ and 
$\tilde{q}_2(\chi^2_{d,\alpha})-\tilde{q}_1(\chi^2_{d,\alpha})> 0$, then under  assumption that $S=\sqrt{n}\Omega^{-1/2}_n(\hat{\theta}_n-\theta)$ and  $U=\sqrt{n}\hat{\Omega}^{-1/2}_n(\hat{\theta}_n-\theta)$ have third-term Edgeworth expansions, 
$\tilde{\mathcal{R}}_{SBP}$ is fourth order-accurate. 
That is $P(\theta\in\tilde{\mathcal{R}}_{SBP})=\alpha+O(n^{-2})$, where

\[\tilde{q}_{1}(\chi^{2}_{d,\alpha})=\frac{2}{d}\chi^{2}_{d,\alpha}a_{1}
+\frac{2}{d}\frac{(\chi^{2}_{d,\alpha})^{2}}{d+2}a_{2}+\frac{2}{d}\frac{(\chi^{2}_{d,\alpha})^{3}}{(d+2)(d+4)}a_{3}.\]

and

\[\tilde{q}_{2}(\chi^{2}_{d,\alpha})=\frac{2}{d}\chi^{2}_{d,\alpha}b_{1}
+\frac{2}{d}\frac{(\chi^{2}_{d,\alpha})^{2}}{d+2}b_{2}+\frac{2}{d}\frac{(\chi^{2}_{d,\alpha})^{3}}{(d+2)(d+4)}b_{3},\]
\end{theorem} 
\noindent
where $a_{i}$ and $b_{i}$ are scalar functions of $\eta_{i_{1},\ldots,i_{k}}=E[(Z_{i_{1}}-\eta_{i_{1}})\cdots(Z_{i_{k}}-\eta_{i_{k}})]$ through the terms that appear in the asymptotic expansions of the cumulants of $S=\sqrt{n}\Omega^{-1/2}_n(\hat{\theta}_n-\theta)$ and  $U=\sqrt{n}\hat{\Omega}^{-1/2}_n(\hat{\theta}_n-\theta)$, respectively. The idea of applying the kernel smoothing technique to a function of data seems new and quite general. The optimal choice of bandwidth matrix $H_n$ given in Theorem \ref{T:Th5} reduces the order of the coverage error of $\tilde{\mathcal{R}}_{BP}$ to $O(n^{-2})$. Depends on the sign of the empirical bandwidth matrix $\hat{H_n}=(n\chi^2_{d,\alpha})^{-1}\{\hat{\tilde{q}}_2(\chi^2_{d,\alpha})-\hat{\tilde{q}}_1(\chi^2_{d,\alpha})\}\hat{\Psi}$, we can construct $\tilde{\mathcal{R}}_{SBP}$ either based on $\hat{\Psi}_n$+ $\hat{H}_n$ or $[1-(n\chi^2_{d,\alpha})^{-1}\{\hat{\tilde{q}}_1(\chi^2_{d,\alpha})-\hat{\tilde{q}}_1(\chi^2_{d,\alpha})\}]\hat{\Psi}$. A similar result is discussed in Section 2.

\subsection{Iterated Bootstrap}
\numberwithin{equation}{section}
In this section we consider the iterated bootstrap method for constructing the bootstrap percentile method ellipsoidal confidence region for $\theta$. Let $S^{**}=\sqrt{n}\hat{\Omega}^{*-1/2}_{n}(\hat{\theta}^{**}_{n}-\hat{\theta}^{*}_{n})$ be the version of $S^{*}$ based on $\mathcal{X}^{**}$, where $\mathcal{X}^{**}$ denotes a generic second level bootstrap sample drawn randomly, with replacement, from $\mathcal{X}^{*}$.  Then we define the theoretical iterated bootstrap  percentile method ellipsoidal confidence region for $\theta$ to be
\[
\tilde{\mathcal{R}}_{RBP}=\{\hat{\theta}_{n}-n^{-1/2}\hat{\Omega}^{1/2}_{n}s:s\in \tilde{\mathcal{S}}_{BP,\alpha+u_n}\}, 
\] 
\noindent  
where $u_n$ satisfies
\[ P[\hat{\theta}_{n}\in \{\hat{\theta}^{**}_{n}-n^{-1/2}\hat{\Omega}^{*1/2}_{n}s:s\in \tilde{\mathcal{S}}^{*}_{BP,\alpha+u_n}\}|\mathcal{X},\mathcal{X}^*]=\alpha.\]
To avoid the double bootstrapping to produce $\tilde{\mathcal{R}}_{RBP}$, we propose  an analytical adjustment to the nominal coverage level. Similar to the mean vector, we can construct \[
\tilde{\mathcal{R}}_{AN}=\{\hat{\theta}_{n}-n^{-1/2}\hat{\Omega}^{1/2}_{n}s:s\in \tilde{\mathcal{S}}_{BP,\alpha+\tilde{u}_{n}}\},\]
where\[ 
\tilde{u}_n=n^{-1} [\hat{\tilde{q}}_{2}(\chi^2_{d,\alpha})-\hat{\tilde{q}}_{1}(\chi^2_{d,\alpha})]g_{d}(\chi^2_{d,\alpha}).\] 

\begin{theorem}\label{T:Th6} Under Assumption of Theorem \ref{T:Th5} , $P(\theta \in \tilde{\mathcal{R}}_{AN})=\alpha+O(n^{-2})$.
\end{theorem}
A discussion for Theorem \ref{T:Th6} is very similar to Theorem \ref{T:Th4}, and is omitted. In the following section we will study the finite sample performance of our proposed  bootstrap percentile method ellipsoidal confidence regions. 

\numberwithin{table}{section}
\section{A Simulation Study}

A simulation study was performed to investigate finite sample performance of $\mathcal{R}_{SBP}$ and $\mathcal{R}_{AN}$.  We compared $\mathcal{R}_{SBP}$ and $\mathcal{R}_{AN}$ with  $\mathcal{R}_{BP}$, $\mathcal{R}_{RBP}$, and  $\mathcal{R}_{BT}$. The performances of different regions were evaluated based on their coverage probabilities and volumes. In this simulation study we consider mean vectors. 

Now we described in detail the setting of our simulation study. Bootstrap regions $\mathcal{R}_{BP}$, $\mathcal{R}_{BT}$, $\mathcal{R}_{SBP}$ and $\mathcal{R}_{AN}$ were constructed using $B=1000$ bootstrap samples and $\mathcal{R}_{RBP}$ was constructed using $C=1000$ inner level bootstrap samples. The coverage probability of various regions were approximated from 10,000 random samples. In our simulation study we considered six different bivariate and trivariate distributions.  See Tables \ref{table:1} and \ref{table:2}.  In these tables
$N(\mu_1,\mu_2,\sigma_1^2,\sigma_2^2,\sigma_{12})]$ denotes a bivariate normal distribution with mean vector $\mu=(\mu_1,\mu_2)'$ and covariance matrix equal to  
\[
\Sigma=
\left[
\begin{matrix}
\sigma_1^2 & \sigma_{12} \\ \sigma_{12} & \sigma_2^2 
\end{matrix}
\right].
\]
\noindent
and $N(\mu_{1},\mu_{2},\mu_{3},\sigma_{1}^{2},\sigma_{2}^{2},\sigma_{12},\sigma_{13},\sigma_{23})$ represents a trivariate normal distribution with mean vector $\mu=(\mu_1,\mu_2,\mu_3)'$ and covariance matrix 
\[
\Sigma=
\left[
\begin{matrix}
\sigma_1^2 & \sigma_{12} & \sigma_{13} \\ 
\sigma_{12} & \sigma_2^2 & \sigma_{23} \\
\sigma_{13} & \sigma_{23} & \sigma_3^2 \\
\end{matrix}
\right].
\]
The form of the normal mixtures used in this study are similar to those of 
Wand and Jones (1993). All these distribution were studies using sample sizes $n=10, 20$.

\begin{table}
\caption{The bivariate normal mixtures used in the simulation. 
The notation used in this table is detailed in Section 4. } 
\begin{center}
\begin{tabular}{cc} 
\hline 
Distribution & Normal Mixture Densities \\
\hline
Normal & $N(0,0,1,1,\frac{1}{2})$ \\ [3pt]
Skewed & 
$\frac{1}{5}N(0,0,1,1,0)+
\frac{1}{5}N(\frac{1}{2},\frac{1}{2},\frac{4}{9},\frac{4}{9},0)$ 
  $+\frac{3}{5}N(\frac{13}{12},\frac{13}{12},\frac{25}{81},\frac{25}{81},0)$ \\ [3pt]
Kurtotic & 
$\frac{2}{3}N(0,0,1,4,1)+\frac{1}{3}N(0,0,\frac{4}{9},\frac{1}{9},-\frac{1}{9})$ \\ [3pt]
Bimodal & 
$\frac{1}{2}N(-1,0,\frac{4}{9},\frac{4}{9},0)+
\frac{1}{2}N(1,0,\frac{4}{9},\frac{4}{9},0)$ \\  [3pt]
Trimodal & 
$\frac{1}{3}N(-\frac{6}{5},0,\frac{9}{25},\frac{9}{25},\frac{63}{250})+
\frac{1}{3}N(\frac{6}{5},0,\frac{9}{25},\frac{9}{25},\frac{63}{250})+
\frac{1}{3}N(0,0,\frac{9}{25},\frac{9}{25},-\frac{63}{250})$ \\ [3pt] 
\hline 
\end{tabular} 
\end{center}
\label{table:1} 
\end{table}

\begin{table}
\begin{center}
\caption{The trivariate normal mixtures used in the simulation. 
The notation used in this table is detailed in Section 4. } 
\begin{tabular}{cc} 
\hline 
Distribution & Normal Mixture Densities\\
\hline
Normal & $N(0,0,0,1,1,1,\frac{3}{10},\frac{2}{5},\frac{1}{2})$ \\ [5pt]
Skewed & 
$\frac{1}{5}N(0,0,0,1,1,1,0,0)+
\frac{1}{5}N(\frac{1}{2},\frac{1}{2},\frac{1}{2},\frac{4}{9},\frac{4}{9},\frac{4}{9},0,0)$ 
\\ [3pt]
& $+\frac{3}{5}N(\frac{13}{12},\frac{13}{12},\frac{13}{12},\frac{25}{81},
\frac{25}{81},\frac{25}{81},0,0)$ \\ [5pt]
Kurtotic & 
$\frac{2}{3}N(0,0,0,1,4,6,1,1,2)+
\frac{1}{3}N(0,0,0,\frac{4}{9},\frac{1}{9},\frac{1}{16},-\frac{1}{9},0,0)$ \\ [5pt]
Bimodal & 
$\frac{1}{3}N(-1,-1,-1,\frac{4}{9},\frac{4}{9},\frac{4}{9},0,0,0)+
\frac{1}{2}N(1,0,0,\frac{4}{9},\frac{4}{9},\frac{4}{9},0,0,0)$ \\  [5pt]
Trimodal & $\frac{1}{3}N(-3,0,0,\frac{9}{25},\frac{9}{25},\frac{9}{25},
\frac{63}{250},0,0)$ \\ [3pt]
 & $+\frac{1}{3}N(3,0,0,\frac{9}{25},\frac{9}{25},\frac{9}{25},\frac{63}{250},
 \frac{63}{250},\frac{63}{250})$ \\ [3pt] 
 & $+\frac{1}{3}N(0,0,0,\frac{9}{25},\frac{9}{25},\frac{9}{25},-\frac{63}{250},0,0)$ \\ 
 [3pt]
\hline 
\end{tabular} 
\end{center} 
\label{table:2}
\end{table}

\begin{table}
\caption{Estimated coverage probabilities of approximate $90\%$ ellipsoidal confidence regions for mean vectors for bivariate distributions.}
\vspace{0.1in}
\centering
\begin{tabular}{ccccccc} 
\hline 
Bivariate distribution   & $n$ &  $\mathcal{R}_{SBP}$ & $\mathcal{R}_{RBP}$ & $\mathcal{R}_{AI}$ & $\mathcal{R}_{BP}$ & $\mathcal{R}_{BT}$\\
\hline   
Independent Normal & 10 & 89.4 & 89.7 & 89.0 & 76.9  & 94.6\\
                   & 20& 90.5 & 90.2 & 89.6   & 84.8  &91.5\\ [7pt]
									
Dependent Normal & 10 &90.4 & 89.3 & 88.9 & 78.2  & 93.4   \\
                 & 20 &89.8 & 90.1 & 89.6 & 84.5  & 91.5   \\ [7pt]
								
Skewed & 10 & 88.3 & 88.3 & 87.8 & 75.4  &92.9    \\
       & 20 & 90.5 & 89.8 & 89.4 & 83.6  &91.6    \\ [7pt]
Kurtotic & 10 & 93.2 & 87.4 &85.2 & 75.0 & 94.5  \\
         & 20 & 90.1 & 90.4 &89.8  & 92.6 & 94.2\\
Bimodal  & 10 & 89.6 & 90.7 &91.3 & 79.0  & 94.1   \\
         & 20 & 89.9& 89.7 & 90.5 &84.9   & 90.0    \\ [7pt]
Trimodal & 10 &90.5 & 90.5 &91.0 &76.6  &93.8   \\ 
         & 20 & 90.3& 89.8 &90.3 &85.6  &91.3  \\ 
\hline  
\end{tabular}  
\label{table:3} 
\end{table}

\begin{table}

\caption{Estimated coverage probabilities of approximate $90\%$ ellipsoidal confidence regions for mean vectors for trivariate distributions.}
\vspace{0.1in}
\centering
\begin{tabular}{ccccccc}				
\hline
Trivariate distribution & $n$ &  $\mathcal{R}_{SBP}$ & $\mathcal{R}_{RBP}$ & $\mathcal{R}_{AI}$ & $\mathcal{R}_{BP}$ & $\mathcal{R}_{BT}$\\
\hline
Independent Normal & 10 & 87.5 & 88.0 &87.4 & 71.3  &97.5     \\
                   & 20 & 89.9& 90.1 &89.7 & 80.6  &91.4   \\ [7pt]
Dependent Normal & 10 & 88.9 & 88.2 & 87.5 & 67.3  & 96.3   \\
                 & 20 & 90.4& 90.0 & 89.4 & 82.1  & 93.1   \\ [7pt]
Skewed & 10 & 85.3& 85.5 & 84.5 & 65.4  &95.6     \\
       & 20 & 89.7& 89.2 & 89.0 & 80.9  &92.1   \\ [7pt]
Kurtotic & 10 & 88.8& 88.3 & 87.5& 70.9   &98.5    \\
         & 20 & 89.3& 89.4 & 89.0 & 80.2   &94.4   \\ [7pt]
Bimodal  & 10 & 87.2& 87.0 & 86.1 & 69.7   & 97.6  \\
         & 20 & 88.9& 88.3 & 87.9& 80.9   & 91.7  \\ [7pt]
Trimodal & 10 & 87.2 & 87.5 & 87.0 &69.0  &96.0  \\ 
         & 20 &  90.7 & 89.8 & 89.6 &82.2  &91.3  \\ 
\hline  
\end{tabular}  
\label{table:4} 
\end{table}

\begin{table}
\caption{Average estimated square radii  of spheres used to compute the bootstrap confidence regions in bivariate case.}
\vspace{0.1in}
\begin{center}
\small
\begin{tabular}{ccccccc} 
\hline 
Bivariate Distribution & $n$ &  $\mathcal{R}_{SBP}$ & $\mathcal{R}_{RBP}$ & $\mathcal{R}_{AI}$ & $\mathcal{R}_{BP}$ & $\mathcal{R}_{BT}$\\   
\hline
Independent Normal & 10 &4.58  & 4.90 & 4.97  & 4.54   &10.44  \\
                   &20  &4.59  & 4.65 & 4.66  & 4.59  &6.13   \\[7pt]
									
Dependent Normal & 10  &4.59  & 4.91 & 4.96  & 4.54 & 10.77   \\
                 &20   & 4.60 & 4.65 & 4.69  & 4.57  &6.11    \\[7pt]
								
Skewed           & 10  & 4.58 & 4.95  &4.97     & 4.52 & 11.91     \\
                  & 20 & 4.59 & 4.66  &4.68     & 4.56 &6.57      \\ [7pt]
									
Kurtotic & 10 & 4.58 & 4.93 & 4.96 & 4.52  & 12.31    \\
         & 20 & 4.60 & 4.67 & 4.70 & 4.59 & 6.30     \\ [7pt]
				
Bimodal  & 10 & 4.59 & 4.92  & 4.96& 4.53 & 10.32   \\
         & 20 &  4.60 &4.65  &4.67 & 4.58 & 6.10     \\ [7pt]
				
Trimodal & 10 & 4.58 & 4.91  &4.94 & 4.53  & 11.03   \\ 
         & 20 & 4.59&  4.66  &4.68 & 4.58  & 6.13    \\ 
\hline
\end{tabular}  
\end{center}
\label{table:5} 
\end{table}
 
\begin{table}
\caption{Average estimated square radii of spheres used to compute the bootstrap confidence regions in trivariate case.}
\vspace{0.1in}
\begin{center}
\small
\begin{tabular}{ccccccc} 
\hline
Trivariate distribution & $n$ & $\mathcal{R}_{SBP}$ & $\mathcal{R}_{RBP}$ & $\mathcal{R}_{AI}$ & $\mathcal{R}_{BP}$ & $\mathcal{R}_{BT}$\\
\hline
Independent Normal & 10 &6.22 &6.85 &7.09 & 6.12  & 26.98   \\
                   &20 &6.23 &6.30 &6.40 &6.21 &9.45\\[7pt]
Dependent Normal & 10 &6.21 &6.84 &7.10 & 6.13 & 10.77 \\
                 & 20 &6.23 &6.30 &6.36 & 6.19 & 9.48 \\[7pt]
Skewed & 10 & 6.22 &6.90 &7.12 & 6.11  & 31.26 \\
       & 20 & 6.23 &6.35 &6.40 & 6.20  & 10.37 \\ [7pt]
Kurtotic & 10 & 6.23 &6.88 &7.10  & 6.13   & 38.92 \\
         & 20 & 6.24 &6.34 &6.37  & 6.21   & 10.23 \\ [7pt]
Bimodal  & 10 & 6.22 &6.88 &6.10 & 6.13  & 26.70 \\
         & 20 & 6.23 &6.30 &6.36 & 6.20   & 9.45 \\ [7pt]
Trimodal & 10 & 6.22 &6.87 &7.12 & 6.11 & 29.03 \\ 
         & 20 & 6.22 &6.31 &6.35 & 6.20  & 9.59 \\
\hline 				
\end{tabular}  
\end{center}
\label{table:6} 
\end{table}

\noindent
The results are presented in Tables \ref{table:3}-\ref{table:6}. We see from Tables \ref{table:3}-\ref{table:4} that $\mathcal{R}_{SBP}$ is much more accurate than $\mathcal{R}_{BP}$ and $\mathcal{R}_{BT}$. This confirms the finite sample gained acquired by the smoothed bootstrap. It should be noted that the coverage probabilities of $\mathcal{R}_{RBP}$ are quite similar to that of $\mathcal{R}_{SBP}$. The coverage probability also demonstrate that $\mathcal{R}_{AI}$ competes closely with $\mathcal{R}_{SBP}$ and $\mathcal{R}_{RBP}$. The degree of improvements by $\mathcal{R}_{SBP}$ and $\mathcal{R}_{AI}$ over  $\mathcal{R}_{BP}$ are remarkable. In general, the coverage probabilities  improve as sample size increases from 10 to 20. Though, for the trivariate distributions the coverage errors increase for all regions when compared to the bivariate cases, especially for $n=10$. 

Another important aspect of confidence region is volume. The results in Tables \ref{table:5} -\ref{table:6} also indicate that the volume of $\mathcal{R}_{SBP}$  is  smaller than that of $\mathcal{R}_{RBP}$, $\mathcal{R}_{AI}$, and $\mathcal{R}_{BT}$ on average. Therefore, overall the $\mathcal{R}_{SBP}$  method outperforms the other regions under consideration.

\section{Discussion}

We have examined the asymptotic effects of using a smoothed bootstrap method in 
conjunction with the bootstrap percentile ellipsoidal confidence for a multivariate smooth function of a mean vector. 
We establish a bandwidth matrix which reduces the asymptotic coverage error of 
the method. By smoothing the bootstrap percentile method we can reduce the asymptotic order of the coverage error of the method to order $O(n^{-2})$.  

In addition to the smoothed bootstrap, we also consider the iterated bootstrap method in constructing a ellipsoidal confidence region. We provide an analytical correction to the nominal coverage probability to avoid the double bootstrapping. Therefore $\mathcal{R}_{AN}$ has the merit computational simplicity.

Our focus on elliptically shaped regions may appear to be unnecessarily restrictive. In fact, the purpose of a nonparametric analysis would be to avoid such restriction. From a theoretical viewpoint such restrictions offer a necessary framework that allows us to study the behavior of the confidence regions using the multivariate Edgeworth expansions. This theory allows us to provide a relatively simple closed form analysis of the problem. In this paper the closed form expression for the bandwidth matrix and the analytical correction term to nominal level are specific only to the ellipsoidal confidence region.

The smoothed bootstrap approach can easily adapted to regions of other shapes. The only difficulty we may face to obtain a closed form expression for the bandwidth matrix for the smoothed bootstrap. However, the data driven smoothed bootstrap approach, despite its high computational cost, is easily applicable for a routine use.

\appendix  
\section{Proofs}

To prove Theorems \ref{T:Th1}-\ref{T:Th4}, we use the following results:

\begin{lemma} \label{L:Le1}  
Let $G_{n}(x)=P(S^{'}S \le x)$, then
\ben\label{eq:(1)}
G_{n}(x)=G_{p}(x)-\frac{2}{np}g_{p}(x)\left[a_{1}x+\frac{a_{2}x^{2}}{p+2}+\frac{a_{3}x^{3}}{(p+2)(p+4)}\right]+O(n^{-2}),
\een 
where
\[a_{1}=\frac{1}{8}\kappa^{(2)}_{3}+\frac{1}{12}\kappa^{(1)}_{3}-\frac{1}{8}\kappa^{(1)}_{4},\] \[a_{2}=\frac{1}{8}\kappa^{(1)}_{4}-\frac{1}{4}\kappa^{(2)}_{3}-\frac{1}{6}\kappa^{(1)}_{3},\]
and
\[
a_{3}=\frac{1}{8}\kappa^{(2)}_{3}+\frac{1}{12}\kappa^{(1)}_{3}.
\]
\end{lemma}

\begin{lemma} \label{L:Le2}
Let $H_{n}(x)=P(U^{'}U \le x)$, then
\ben
H_{n}(x)= G_{p}(x)-\frac{2}{np}g_{p}(x)\left[b_{1}x+\frac{b_{2}x^{2}}{p+2}+
\frac{b_{3}x^{3}}{(p+2)(p+4)}\right]+O(n^{-2}),
\een
where
\[
b_{1}=\frac{p(p+2)}{4}+\frac{1}{2}\kappa^{(1)}_{4}- \frac{1}{6}\kappa^{(1)}_{3},
\]
\[
b_{2}= \frac{p(p+2)}{4}+\frac{1}{3}\kappa^{(1)}_{3}- \frac{1}{4}\kappa^{(1)}_{4},
\]
and
\[
b_{3}=\frac{1}{3}\kappa^{(1)}_{3}+\frac{1}{2}\kappa^{(2)}_{3}.        
\]
\end{lemma}

\begin{lemma} \label{L:Le3} 
Let $r^{2}_{S}$ and $r^{2}_{U}$ be the square radii of the spheres $\mathcal{S}_S$ and $\mathcal{S}_U$ (see the discussion of Theorem \ref{T:Th1}). Then
\[r^{2}_{S}=\chi^{2}_{p,\alpha}+n^{-1}q_{1}(\chi^{2}_{p,\alpha})+O(n^{-2}),\] and

\[r^{2}_{U}=\chi^{2}_{p,\alpha}+n^{-1}q_{2}(\chi^{2}_{p,\alpha})+O(n^{-2}).\] 
\end{lemma}

Let $r^{2}_{BP}$ be the square radii of the sphere $\mathcal{S}_{BP}$ and $r^{2}_{BP}$ has same expansion as $r^{2}_{S}$ by replacing $q_{1}(\chi^{2}_{p,\alpha})$ by its sample version.

\begin{lemma} \label{L:Le4} 
Let $r^{2}_{SBP}$ and $r^{2}_{BP}$ be the square radii of the spheres $\mathcal{S}_{SBP}$ and $\mathcal{S}_{BP}$, where $\mathcal{S}_{SBP}$ and $\mathcal{S}_{BP}$ correspond to the ellipsoidal regions $\mathcal{R}_{SBP}$ and $\mathcal{R}_{BP}$. Then
\[r^{2}_{SBP}=r^{2}_{BP}+O_{p}(n^{-\min(k+1, 2)}),\]
where $k>0$ is such that $H_{n}=O(n^{-k})$. 
\end{lemma}

\noindent
For the interest of space, we only provided the proofs of the theorems in this paper. Proofs of the lemmas 1--4, are provided in Supplement A.

\textsc{\textbf{Proof of Theorem} \ref{T:Th1}}. 
The coverage probability of percentile ellipsoidal confidence region for $\theta$ is 
\bea\label{eq:A1}
P(\theta\in{\mathcal{R}}_{BP})=P(\theta\in\hat{\theta}_{n}-n^{-1/2}\hat{\Sigma}^{1/2}_{n}{\mathcal{S}}_{BP})=P(U'U\leq r^{2}_{BP})\nonumber\\
= P(U'U\leq r^{2}_{S}r^{-2}_{S}r^{2}_{BP})\nonumber\\
=P(r^{2}_{S}r^{-2}_{BP}U'U\leq r^{2}_{S})
\eea
From Lemma \ref{L:Le3} we have that 
\bea\label{eq:A2}
 r^{2}_{S}r^{-2}_{BP}&=&[\chi^2_{p,\alpha}+n^{-1}q_1(\chi^2_{p,\alpha})+O(n^{-2})][\chi^2_{p,\alpha}+n^{-1}\hat{q}_1(\chi^2_{p,\alpha})+O_{p}(n^{-2})]^{-1}\nonumber\\
&=&[1+(n\chi^2_{p,\alpha})^{-1}q_1(\chi^2_{p,\alpha})+O(n^{-2})][1-(n\chi^2_{p,\alpha})^{-1}\hat{q}_1(\chi^2_{p,\alpha})+O(n^{-2})]\nonumber\\
&=&1+(n\chi^2_{p,\alpha})^{-1}[q_1(\chi^2_{p,\alpha})-\hat{q}_1(\chi^2_{p,\alpha})]+O_{p}(n^{-2})\nonumber\\
&=&1+O_{p}(n^{-3/2}),\nonumber\\
\eea
last line follow from the fact that $\hat{q}_1(\chi^2_{p,\alpha})=q_1(\chi^2_{p,\alpha})+O_{p}(n^{-1/2})$. Therefore, Equations (\ref{eq:A1}) and (\ref{eq:A2}) yield 
\bea\label{eq:A3}
P(\theta\in\mathcal{R}_{BP})=
P[(1+O_{p}(n^{-3/2}))U^{'}U\leq r^{2}_{S}]\nonumber\\
=P[(1+\Delta_{n})^{1/2}U\in \mathcal{S}_{S}],\nonumber\\ 
\eea
\noindent
where $\Delta_{n}=O_{p}(n^{-3/2})$. The Edgeworth expansion of the distribution of $(1+\Delta_{n})^{1/2}U$ then implies that
\begin{multline*}
P\left(\theta\in{\mathcal{R}}_{BP}\right)=\\
\int_{\mathcal{S}_{S}}[1+n^{-1/2}t_{1}(x)+n^{-1}t_{2}(x)+n^{-3/2}t_{3}(x)]
\phi_{p}(x)dx + O(n^{-2}). 
\end{multline*}
Since $U$ and $(1+\Delta_{n})^{1/2}U$ differ only in terms of order 
$O_{p}(n^{-3/2})$ it follows that $t_j=p_j$ for $j=1,2$, where 
$p_j$ is the polynomial that appears in the Edgeworth expansion for the distribution 
of $U$. Therefore 
\begin{equation}\label{eq:A4}
P\left(\theta\in{\mathcal{R}}_{BP}\right)=P(U\in {\mathcal{S}}_{S})+O(n^{-2}).
\end{equation}
\noindent
From Equation (\ref{eq:A4}) we have that 
\begin{eqnarray}\label{eq:A5}
P\left(\theta\in{\mathcal{R}}_{BP}\right)= 
P(U\in {\mathcal{S}}_{U})+O(n^{-2})\nonumber\\
=P[U'U\leq \chi^{2}_{p,\alpha}+n^{-1}q_{1}(\chi^{2}_{p,\alpha})+O(n^{-2})]
 +O(n^{-2}).\nonumber\\
\end{eqnarray}
\noindent
Applying Lemma \ref{L:Le2} to Equation (\ref{eq:A5}) yields
\[
P\left(\theta\in{\mathcal{R}}_{BP}\right)=
\alpha+n^{-1}[q_{1}(\chi^{2}_{p,\alpha})-q_{2}(\chi^{2}_{p,\alpha})]
g_{p}(\chi^{2}_{p,\alpha})+O(n^{-2}).
\]

\textsc{\textbf{Proof of Theorem \ref{T:Th2}}}. 
The coverage probability of the confidence region 
$\mathcal{R}_{SBP}$ is given by  
\bea\label{eq:A6}
P(\theta\in\mathcal{R}_{SBP})= 
P[\sqrt{n}\tilde{\Sigma}^{-1/2}_n(\hat{\theta}_n-\theta)\in\mathcal{S}_{SBP}]
\nonumber \\
= P[n(\hat{\theta}_n-\theta)'\tilde{\Sigma}^{-1}_n
 (\hat{\theta}_n-\theta)\leq r^2_{SBP}]\nonumber,\\
\end{eqnarray}
where $r_{SBP}$ is the radius of the sphere $\mathcal{S}_{SBP}$. It can be shown that 
$\tilde{\Sigma}^{-1}_n=\hat{\Sigma}^{-1}_n-\hat{\Sigma}^{-1}_n H_n\hat{\Sigma}^{-1}_{n}+O_{p}(n^{-2k})$. See Seber (2008, Section 15.1). From Equation (\ref{eq:A6}) it follows that

\bea\label{eq:A7}
P(\theta\in\mathcal{R}_{SBP})=  P\{n(\hat{\theta}_n-\theta)'[\hat{\Sigma}^{-1}_n-\hat{\Sigma}^{-1}_n H_n\hat{\Sigma}^{-1}_{n}+O_{p}(n^{-2k})](\hat{\theta}_n-\theta)\leq r^2_{SBP}\}\nonumber,\\
= P\{U'[I-\hat{\Sigma}^{-1/2}_n H_n\hat{\Sigma}^{-1/2}_{n}+O_{p}(n^{-2k})]U\leq r^2_{SBP}\}\nonumber,\\
= P\{U'[r^{-2}_{SBP}r^2_{U}(I-\hat{\Sigma}^{-1/2}_n H_n\hat{\Sigma}^{-1/2}_{n}+O_{p}(n^{-2k}))]U\leq r^2_{U}\}\nonumber,\\
\end{eqnarray}
Using Lemma \ref{L:Le3}, Lemma\ref{L:Le4} and fact  $\hat{q}_1(\chi^2_{p,\alpha})=q_1(\chi^2_{p,\alpha})+O_{p}(n^{-1/2})$, we have 
\bea \label{eq:A8}
r^{-2}_{SBP}r^{2}_u &=&  
[r^2_{BP}+O_p(n^{-\min(k+1,2)})]^{-1}[\chi^2_{p,\alpha}+n^{-1}q_2
(\chi^2_{p,\alpha})+O(n^{-2})] \nonumber\\
&=& [\chi^2_{p,\alpha}+n^{-1}\hat{q}_1
(\chi^2_{p,\alpha})+O_p(n^{-\min(k+1,2)})]^{-1}[\chi^2_{p,\alpha}+n^{-1}q_2
(\chi^2_{p,\alpha})+O(n^{-2})] \nonumber\\
&=&1+(n\chi^{2}_{p,\alpha})^{-1}[q_{2}
 (\chi^{2}_{p,\alpha})-q_{1}(\chi^{2}_{p,\alpha})]+O_{p}(n^{-\min(k+1,3/2)}). 
\eea
and combing Equation (\ref{eq:A8}) with $[I-\hat{\Sigma}^{-1/2}_n H_n\hat{\Sigma}^{-1/2}_{n}+O_{p}(n^{-2k})]$ yield
\bea \label{eq:A9}
r^{-2}_{SBP}r^{2}_u [I-\hat{\Sigma}^{-1/2}_n H_n\hat{\Sigma}^{-1/2}_{n}+O_{p}(n^{-2k})]  
=\{1+(n\chi^{2}_{p,\alpha})^{-1}[q_{2}
 (\chi^{2}_{p,\alpha})-q_{1}(\chi^{2}_{p,\alpha})]\nonumber\\
+O_{p}(n^{-\min(k+1,3/2)})\}[I-\hat{\Sigma}^{-1/2}_n H_n\hat{\Sigma}^{-1/2}_{n}+O_{p}(n^{-2k})]\nonumber\\
= \{1+(n\chi^{2}_{p,\alpha})^{-1}[q_{2}(\chi^{2}_{p,\alpha})-q_{1}(\chi^{2}_{p,\alpha})]\}[I-\hat{\Sigma}^{-1/2}_n H_n\hat{\Sigma}^{-1/2}_{n}]+O_{p}(n^{-\min(k+1,3/2,2k)})\nonumber\\
\eea
 Equation (\ref{eq:A9}) and the fact $\hat{\Sigma}^{-1/2}_{n}=\Sigma^{-1/2}+O_{p}(n^{-1/2})$  yield  
\begin{multline}\label{eq:A10}
U'\{1-(n\chi^{2}_{p,\alpha})^{-1}[q_1(\chi^{2}_{p,\alpha})-
q_{2}(\chi^{2}_{p,\alpha})+O_{p}(n^{-\min(k+1,3/2,2k)}][I-\hat{\Sigma}^{-1/2}_{n}H_{n}
\hat{\Sigma}^{-1/2}_{n}]\}U \\
 = U'\{I+(n\chi^{2}_{p,\alpha})^{-1}[q_{2}(\chi^{2}_{p,\alpha})-q_1
 (\chi^{2}_{p,\alpha})]I \\
-\Sigma^{-1/2}H_n\Sigma^{-1/2}+O_{p}(n^{-\min(k+1,3/2,2k)})\}U.
\end{multline}
The bandwidth matrix $H_n$ can be chosen to eliminate the term of order  
$O(n^{-1})$ in Equation (\ref{eq:A10}). 
Therefore, we can take  $H_n=(n\chi^2_{p,\alpha})^{-1}[q_2(\chi^2_{p,\alpha})-q_1(\chi^2_{p,\alpha})]\Sigma$, provided  $q_{2}(\chi^{2}_{p,\alpha})-q_{1}(\chi^{2}_{p,\alpha})>0$. Because $q_{2}(\chi^{2}_{p,\alpha})-q_{1}(\chi^{2}_{p,\alpha})$ 
does not depend on $n$, the order of $H_{n}$ is $O(n^{-1})$. Hence, with this choice of $H_{n}$,  Equations (\ref{eq:A7}) and (\ref{eq:A10}) imply 
\begin{equation}\label{eq:A11}
P(\theta\in\mathcal{R}_{SBP})=
P[U'\{1+O_{p}(n^{-3/2})\}U \leq r^{2}_{u}].
\end{equation} 
The remainder of the proof follows using the same general arguments as Theorem \ref{T:Th1}. 

\textsc{\textbf{Proof of Theorem} \ref{T:Th3}}. 
The coverage probability of the confidence region $\hat{\mathcal{R}}_{SBP}$ 
is given by  
\begin{multline}\label{eq:A12}
P(\theta\in\hat{\mathcal{R}}_{SBP}) = 
P[\sqrt{n}(\hat{\Sigma}_{n}+\hat{H}_{n})^{-1/2}
(\hat{\theta}_{n}-\theta)\in\hat{\mathcal{S}}_{SBP}] \\
= P\{(1+(n\chi^{2}_{p,\alpha})^{-1}[\hat{q}_{2}(\chi^{2}_{p,\alpha})-
\hat{q}_{1}(\chi^{2}_{p,\alpha})])^{-1}n(\hat{\theta}_{n}-\theta)^{'}\hat{\Sigma}^{-1}_{n}(\hat{\theta}_{n}-\theta)\leq\hat{r}^{2}_{SBP}\}\\
=P\{n(\hat{\theta}_{n}-\theta)^{'}\hat{\Sigma}^{-1}_{n}(\hat{\theta}_{n}-\theta)\leq(1+
(n\chi^{2}_{p,\alpha})^{-1}[\hat{q}_{2}(\chi^{2}_{p,\alpha})-
\hat{q}_{1}(\chi^{2}_{p,\alpha})])\hat{r}^{2}_{SBP}\}, 
\end{multline}
\noindent
where $\hat{r}_{SBP}$ is radius of the sphere $\hat{\mathcal{S}}_{SBP}$. Using Lemma \ref{L:Le4} we have
\begin{multline}\label{eq:A13}
[1+(n\chi^2_{p,\alpha})^{-1}
\{\hat{q}_2(\chi^2_{p,\alpha})-\hat{q}_1
(\chi^2_{p,\alpha})\}]\hat{r}^2_{SBP} = \\ 
r^{2}_{BP}+[\chi^{2}_{p,\alpha}+
n^{-1}\hat{q}_{1}(\chi^{2}_{p,\alpha})] 
(n\chi^{2}_{p,\alpha})^{-1}[\hat{q}_{2}(\chi^{2}_{p,\alpha}) 
 -\hat{q}_{1}(\chi^{2}_{p,\alpha})]+O_{p}(n^{-2}) \\
= \chi^2_{p,\alpha}+n^{-1}\hat{q}_{1}(\chi^2_{p,\alpha})+
n^{-1}\hat{q}_2(\chi^2_{p,\alpha})-
n^{-1}\hat{q}_1(\chi^2_{p,\alpha}) 
  +O_{p}(n^{-2})  \\
=\chi^{2}_{p,\alpha}+n^{-1}\hat{q}_{2}(\chi^{2}_{p,\alpha})+O_{p}(n^{-2})
=r^{2}_{BT},
\end{multline}
which is the square radius of $\mathcal{S}_{BT}$, corresponding to 
the percentile-$t$ method. Hence Theorem \ref{T:Th3} follows from 
Equations (\ref{eq:A12}) and (\ref{eq:A13}).

\textsc{\textbf{Proof of Theorem} \ref{T:Th4}}.
The coverage probability of $\mathcal{R}_{AI}$ is given by
\bea \label{eq:A14}
P\{\theta \in \mathcal{R}_{AI}\}= P\{\theta\in \hat{\theta}_{n}-n^{-1/2}\hat{\Sigma}^{1/2}_{n}\mathcal{S}_{BP,\alpha+\tilde{u}_{n}}\}\nonumber\\
= P(U\leq r^{2}_{BP, \alpha+\tilde{u}_{n}}).
\eea
\noindent
Again,
\bea \label{eq:A15}
r^{2}_{BP,\alpha+\tilde{u}_{n}}&=&\chi^2_{p,\alpha+\tilde{u}_{n}}+n^{-1}\hat{q}_{1}(\chi^2_{p,\tilde{u}_{n}})+O_{p}(n^{-2})\nonumber\\
&=&\chi^2_{p, \alpha}+\tilde{u}_{n} [g_{p}(\chi^2_{p})]^{-1}+n^{-1}\hat{q}_{1}(\chi^2_{p,\alpha}+O_{p}(n^{-1}))+O_{p}(n^{-2})\nonumber\\
&=&\chi^2_{p,\alpha}+n^{-1} [\hat{q}_{2}(\chi^2_{p,\alpha})-\hat{q}_{1}(\chi^2_{p,\alpha})]g_{p}(\chi^2_{p,\alpha}) [g_{p}(\chi^2_{p})]^{-1}+n^{-1}\hat{q}_{1}(\chi^2_{p,\alpha})+O_{p}(n^{-2})\nonumber\\
&=&\chi^2_{p,\alpha}+n^{-1} [\hat{q}_{2}(\chi^2_{p,\alpha})-\hat{q}_{1}(\chi^2_{p,\alpha})]+ n^{-1}\hat{q}_{1}(\chi^2_{p,\alpha})+O_{p}(n^{-2})\nonumber\\
&=&\chi^2_{p,\alpha}+n^{-1} \hat{q}_{2}(\chi^2_{p,\alpha})+O_{p}(n^{-2})\nonumber\\
&=& r^{2}_{BT},\nonumber\\
\eea
which is the square radius of $\mathcal{S}_{BT}$, corresponding to the percentile-$t$ method. Hence Theorem \ref{T:Th4} follows from Equations (\ref{eq:A14}) and (\ref{eq:A15}).\\

\noindent
The proofs of the results in Section 3 follow the arguments of the results in Section 2  with only minor changes due to the increased complexity of the notation involved, and are omitted.


\newpage

\begin{thebibliography}{9}

\bibitem{}
\textsc{Barndorff-Nielsen, O.E.} and \textsc{Cox, D.R.} (1989). 
\textit{Asymptotic Techniques for Use in Statistics}. 
Chapman and Hall, London.

\bibitem{}
\textsc{Beran, R.} (1987). 
Pre-pivoting to reduce level error of confidence sets. 
\textit{Biometrika}, 74, 457--468. 


\bibitem{}
\textsc{Efron, B.} (1979). 
Bootstrap methods: Another look at the jackknife. 
\textit{The Annals of Statistics}, \textbf{7}, 1--26.

\bibitem{}
\textsc{Efron, B.} (1981). 
Nonparametric standard errors and confidence intervals (with discussion).  
\textit{Canadian Journal of Statistics}, \textbf{9}, 139--172.

\bibitem{}
\textsc{Efron, B.} (1982). 
\textit{The Jackknife, the bootstrap, and other Resampling Plans}. 
\textit{SIAM, Philadelphia.}

\bibitem{}
\textsc{Efron, B.} (1987). 
Better bootstrap confidence intervals. 
\textit{Journal of American Statistical Association}, \textbf{82}, 171--200.

\bibitem{}
\textsc{Fujikoshi, Y.} (1997). 
An asymptotic expansion for the distribution of Hotelling's $T^{2}$-statistics  
under nonnormality. 
\textit{Journal Of Multivariate Analysis}, \textbf{61}, 187--193.



\bibitem{}
\textsc{Guerra, R.}, \textsc{Polansky, A.M.} and \textsc{Schucany, W.R.} (1997). 
Smoothed bootstrap confidence intervals with discrete data. 
\textit{Computational Statistics and Data Analysis}, \textbf{26}, 163--176.


\bibitem{}
\textsc{Hall, P.} (1988). 
Theoretical comparison of bootstrap confidence intervals.  
\textit{The Annals of Statistics}, \textbf{16}, 927--985.

\bibitem{}
\textsc{Hall, P.} (1992). 
\textit{The Bootstrap and Edgeworth Expansion}. 
Springer, New York. 


\bibitem{}
\textsc{Isogai, T.} (1983). 
On measures of multivariate skewness and kurtosis. 
\textit{Mathematica Japanica}, \textbf{28}, 251--261.


\bibitem{}
\textsc{Loh, W.-Y.} (1987). 
Calibrating confidence coefficients. 
\textit{Journal of American Statistical Association}, \textbf{82}, 155--162.

\bibitem{}
\textsc{Mardia, K.V.} (1970). 
Measures of multivariate skewness and kurtosis with application. 
\textit{Biometrika}, \textbf{57}, 519--530.


\bibitem{}
\textsc{Polansky, A. M.} (2001). 
Bandwidth selection for the smoothed bootstrap percentile method.
\textit{Computational Statistics and Data Analysis}, \textbf{36}, 333--349.

\bibitem{}
\textsc{Polansky, A. M.} and \textsc{Schucany, W.R.} (1997). 
Kernel smoothing to improve bootstrap confidence intervals. 
\textit{Journal of the Royal Statistical Society: Series B}, \textbf{59}, 821--838.

\bibitem{}
\textsc{Seber, G. A. F.} (2007). 
\textit{A Matrix Handbook for Statisticians}. 
John Wiley and Sons, New York. 

\bibitem{}
\textsc{Wand, M.P.} and \textsc{Jones, M.C.} (1995). 
\textit{Kernel Smoothing}. 
Chapman and Hall, London. 


\end{thebibliography}
\end{document}